\documentclass[usegraphicx]{mn2e}
\usepackage{subfigure}
\usepackage{epsf}
\usepackage{amsmath}
\usepackage{amssymb}

\newcommand {\apgt} {\ {\raise-.5ex\hbox{$\buildrel>\over\sim$}}\ }
\newcommand {\aplt} {\ {\raise-.5ex\hbox{$\buildrel<\over\sim$}}\ }

\title[Diagnosing the accretion flow in ULXs using soft X-ray atomic features]
{Diagnosing the accretion flow in ultraluminous X-ray sources using soft X-ray atomic features}

\author[M.Middleton et al.]
{Matthew J. Middleton$^{1}$, Dominic J. Walton$^{2, 3}$, Andrew Fabian$^{1}$, Timothy P. Roberts$^{4} $\newauthor Lucy Heil$^{5}$, Ciro Pinto$^{1}$, Gemma Anderson$^{6}$ and Andrew Sutton$^{7}$\\
\\
1. Institute of Astronomy, Madingley Road, Cambridge CB3 OHA\\
2. NASA Jet Propulsion Laboratory, 4800 Oak Grove Dr, Pasadena, CA 91109, USA\\
3. California Institute of Technology, 1200 East California Boulevard, Pasadena, California 91125, USA\\
4. Department of Physics, University of Durham, Durham DH1 3LE\\
5. Astronomical Institute Anton Pannekoek, Science Park 904,1098 XH, Amsterdam, Netherlands\\
6. Department of Physics, Oxford University, Denys Wilkinson Building, Keble Road, Oxford OX1 3RH\\
7. Astrophysics Office, NASA Marshall Space Flight Center, ZP12, Huntsville, Al 35812, USA
}

\pagerange{\pageref{firstpage}--\pageref{lastpage}} \pubyear{2011}
\long\def\symbolfootnote[#1]#2{\begingroup\def\thefootnote{\fnsymbol{footnote}}\footnote[#1]{#2}\endgroup} 
\def\ga{\mathrel{\hbox{\rlap{\hbox{\lower4pt\hbox{$\sim$}}}{\raise2pt\hbox{$>$}}
}}}

\begin{document}

\topmargin = -0.5cm

\maketitle

\label{firstpage}

\begin{abstract}
The lack of unambiguous detections of atomic features in the X-ray spectra of ultraluminous X-ray sources (ULXs) has proven a hindrance in diagnosing the nature of the accretion flow. The possible association of spectral residuals at soft energies with atomic features seen in absorption and/or emission and potentially broadened by velocity dispersion could therefore hold the key to understanding much about these enigmatic sources. Here we show for the first time that such residuals are seen in several sources and appear extremely similar in shape, implying a common origin. Via simple arguments we assert that emission from extreme colliding winds, absorption in a shell of material associated with the ULX nebula and thermal plasma emission associated with star formation are all highly unlikely to provide an origin. Whilst CCD spectra lack the energy resolution necessary to directly determine the nature of the features (i.e. formed of a complex of narrow lines or intrinsically broad), studying the evolution of the residuals with underlying spectral shape allows for an important, indirect test for their origin. The ULX NGC 1313 X-1 provides the best opportunity to perform such a test due to the dynamic range in spectral hardness provided by archival observations. We show through highly simplified spectral modelling that the strength of the features (in either absorption or emission) appears to anti-correlate with spectral hardness, which would rule out an origin via reflection of a primary continuum and instead supports a picture of atomic transitions in a wind or nearby material associated with such an outflow.

\end{abstract}

\section{Introduction}

Although observed for many decades, the nature of ULXs has long been shrouded in mystery. The faintness of their optical counterparts generally inhibits dynamical mass determination (although see Lui et al. 2013 and Motch et al. 2014) thus failing to resolve degenerate scenarios for the nature of the compact object driving the extraction of energy via a luminous accretion flow. The gradual increase in available data quality in recent years (as well as the broader bandpass provided by {\it NuSTAR}: Bachetti et al. 2013; Walton et al. 2013a; 2014; 2015; Rana et al. 2015; Mukherjee et al. 2015) has facilitated ever more detailed studies of ULXs, and dedicated observing campaigns have yielded important insights. Notably, the detection of ballistic jets in a source at the faint end of the ULX regime (Middleton et al. 2013; Middleton, Miller-Jones \& Fender 2014) implies that the majority of those with `broadened disc spectra' (typically at $<$ 3 $\times$10$^{39}$ erg s$^{-1}$: Sutton, Roberts \& Middleton 2013) are probably associated with `normal' modes of accretion onto stellar mass black holes (with M $<$ 100 M$_{\odot}$). These sources have now been observed to undergo the typical spectral evolution seen in Galactic black hole binaries (BHBs, e.g. Fender, Belloni \& Gallo 2004) and when brightest (i.e. when they appear as a ULX), enter an unusual spectral state probably as a result of physical changes in the inner regions of the accretion flow due to radiation and/or magnetic pressure (Middleton et al. 2012; Straub, Done \& Middleton 2013). ULXs which are even brighter show distinctly different X-ray spectra (e.g. Gladstone et al. 2009; Sutton, Roberts \& Middleton 2013; Middleton et al. 2015, hereafter M15) and may be associated with accretion (at rates $\le$ the Eddington rate) onto intermediate mass black holes (IMBHs) with M$_{\rm BH} >$ 100s M$_{\odot}$ (Colbert \& Mushotzky 1999; Pasham, Strohmayer \& Mushotzky 2014), accretion onto low mass SMBHs in stripped dwarf galaxies (Farrell et al. 2009; Webb et al. 2011) or mass transfer at super-Eddington rates and a super-critical accretion flow onto neutron stars (King 2001; Bachetti et al. 2014) or stellar mass black holes (King 2009; Poutanen et al. 2007; Middleton et al. 2011; M15), or  - more likely - a combination of all of these and a heterogenous population. Theoretical formation mechanisms would argue against a large proportion of this more luminous population being associated with IMBHs (King 2004). Indeed, there is mounting evidence for a wider association with high (possibly super-critical) accretion rates onto stellar remnant black holes. A recent dynamical study of one such ULX has shown that the characteristic ultraluminous X-ray spectrum (Gladstone et al. 2009; Sutton et al. 2013) is likely associated with high (super-Eddington) accretion rates onto a compact object of stellar mass (Motch et al. 2014) with the detection of ballistic jets in Ho II X-1 providing independent support for high rates of accretion onto a compact object $>$ 25 M$_{\odot}$ (Cseh et al. 2014). Such identifications have only been possible for a very small number of sources and so the nature of the accretion flow and compact object in the vast majority of ULXs remains somewhat ambiguous. 

Atomic features seen in emission or absorption and imprinted onto the X-ray continuum provide powerful diagnostics of the accretion flow around a black hole, including information about the strength of winds driven from the disc (Ponti et al. 2012) and the nature of optically thick material seen in reflection (Fabian et al. 1989). Due to the high relative abundance of Fe coupled with the high fluorescent yield (which scales as $\rm{Z}^{4}$), the bound-free transitions associated with neutral through to ionised species of this element typically produce the strongest atomic features in the spectrum. The hardest ULXs are those where we expect such features to be most readily detectable due to the substantial amount of hard ($>$ 6~keV) flux needed to excite the relevant transitions. However, to-date, no such features have been unambiguously detected, with stringent upper limits placed through analysis of long exposure {\it Suzaku} observations (Walton et al. 2013b) which imply either a very low Fe abundance, a very low reflected fraction or a highly ionised scattering plasma. The latter is a general prediction of the super-critical model of ULXs (e.g. Poutanen et al. 2007) where very high (super-Eddington) mass transfer rates through an accretion disc around a stellar mass black hole (or equally a neutron star: King 2001), leads to the production of a large scale-height inflow and initially optically thick wind extending from the spherization radius down to the innermost disc edge (Shakura \& Sunyaev 1973; Poutanen et al. 2007). The inflow and wind then traps the copious amounts of radiation from the innermost regions which ionises the inner face of the wind cone radially out to 100s of R$_{\rm g}$ (Middleton et al. 2015). In this same model, the brightest and hardest ULXs are those seen face-on (made brighter due to geometrical beaming: King et al. 2009), providing a natural explanation for the lack of features in these sources.
 
Spectral residuals to the best-fitting continuum models have been identified at soft energies ($<$ 2 keV) in multiple sources using {\it XMM-Newton} data (e.g. Stobbart, Roberts \& Wilms 2006). These cannot be associated with absorption and emission by the nebulae surrounding some ULXs as the densities of these media are expected to be extremely low (typically $\lesssim$1 cm$^{-3}$: Osterbrock 1989; Roberts et al. 2003). An origin in thermal plasma emission associated with high rates of star formation has been suggested, however, Middleton et al. (2014; hereafter M14) argue that the luminosity in such a component is too high (typically 10$^{38-39}$ erg s$^{-1}$) to be reconciled with the expected thermal continuum (bremmstrahlung) and collisionally excited line emission local to ULXs (see also Sutton et al. in prep). M14 instead proposed that the soft residuals to the best-fitting continuum model in two of the softest and brightest ULXs could be reconciled with atomic absorption features, potentially broadened and analogous to those UV absorption lines seen in broad absorption line Quasars (BALQSOs, e.g. Hazard et al. 1984). M14 argue that such features are a corollary of the super-critical ULX model as expanded, optically thin material at large distances should imprint absorption (and emission) features which, when viewed at moderate inclinations (i.e. into the wind) should be broadened by the velocity dispersion along the line-of-sight. 

Whilst there may be compelling arguments for the presence of winds in many of the brightest ULXs, the soft residuals in their spectra cannot yet be unambiguously associated with an outflow as atomic features could also be imprinted by more distant material (presumably expelled via a disc-wind and/or jet) photo-ionised by the ULX, reflection of a primary continuum by an ionised or Fe deficient disc atmosphere, or emission by collisionally excited material, most likely in a disc wind interacting with that from the stellar companion. In addition, whilst CCD instruments have the sensitivity to allow the ULX continuum to be described and allow limits to be placed on the presence of relatively narrow features, they lack the energy resolution necessary to distinguish between a complex of narrow lines (as, for example, may be found originating in more distant, slowly moving material) or intrinsically broad lines (which would likely imply a line-of-sight velocity dispersion and an outflow). Whilst dispersion gratings can provide the requisite energy resolution, they typically lack the sensitivity to reliably study such faint sources (although see Pinto, Middleton \& Fabian in prep). However, we can already attempt to rule out certain origins for the residuals via simple energetic arguments. Typical luminosities expected from colliding wind binaries are only very modest (10$^{32}$ - 10$^{34}$ erg s$^{-1}$: Oskinova 2005 and Mauerhan et al. 2010), whilst crude limits on the expected luminosity of bremsstrahlung cooling from the interaction of a disc wind and stellar wind can be determined for an assumed outflow velocity ($v_{\rm out}$), mass loss rate ($\dot{m}_{\rm out}$) and binary separation ($a_{\rm 12}$, e.g. Cooke, Fabian \& Pringle 1978). By analogy with the Galactic super-critical source, SS433 (see the review of Fabrika et al. 2004) and the outflow velocity expected for radiation pressure driven winds (e.g. King 2010), we have what may be considered broadly reasonable values for ULXs {\it if} super-critical accretors with powerful winds of $v_{\rm out}$ = 0.1c, $\dot{m}_{\rm out}$ = 10$^{-4}$ M$_{\odot}$ yr$^{-1}$ and $a_{\rm 12}$ = 10$^{12}$ cm. It is then possible to see that, whilst the temperature of the emission is consistent with plasma models used to explain the residuals (i.e. 1-2~keV), the luminosity is expected to be $\sim$10$^{37}$ erg s$^{-1}$, one or two orders of magnitude below that needed. Whilst shock heating of the wind itself due to internal collisions between elements (e.g. Takeuchi, Ohsuga \& Mineshige 2013) could be a further source of line emission, if the optical depth is high, this radiation should be mostly thermalised before escaping. The (most obvious) remaining options for the residuals are atomic features seen in emission and absorption in a wind and/or in more distant material, or via reflection. As each of these possible solutions relies on geometry and illumination, by studying how the strength of the features change with the continuum and by making comparisons to predictions, we can attempt to constrain their origin.

\vspace{-0.6cm}
\section{Soft residuals in bright ULXs}

In order to study the spectral residuals, M14 selected the two bright and soft ULXs, NGC 5408 X-1 and NGC 6946 X-1, as these show the strongest residuals to the best-fitting continuum model in archival {\it XMM-Newton} data. However, identifying the {\it systematic} presence of residuals across the larger population is important as it would suggest a ubiquity in their nature and origin. M15 study a sample of ULXs (again using archival {\it XMM-Newton} data) of which Ho II X-1, Ho IX X-1, NGC 5204 X-1, NGC 55 ULX-1, NGC 1313 X-1 and NGC 1313 X-2 have generally high S/N in the soft band of interest and cover a range in spectral `hardness'; these are therefore optimal sources for searching for spectral residuals of the form seen in NGC 5408 X-1 and NGC 6946 X-1. Details of the respective observations are given in M15 where we follow the standard extraction procedure for the spectral products. We also include any newly available {\it XMM-Newton} observations of the sources (namely NGC 1313 X-1/X-2, Ho IX X-1 and Ho II X-1) which are discussed in Bachetti et al. (2013); Walton et al. (2014) and Walton et al. (2015) respectively (and for which we use the latest calibration files).

 \begin{figure}
\begin{center}
\begin{tabular}{l}
 \epsfxsize=8cm \epsfbox{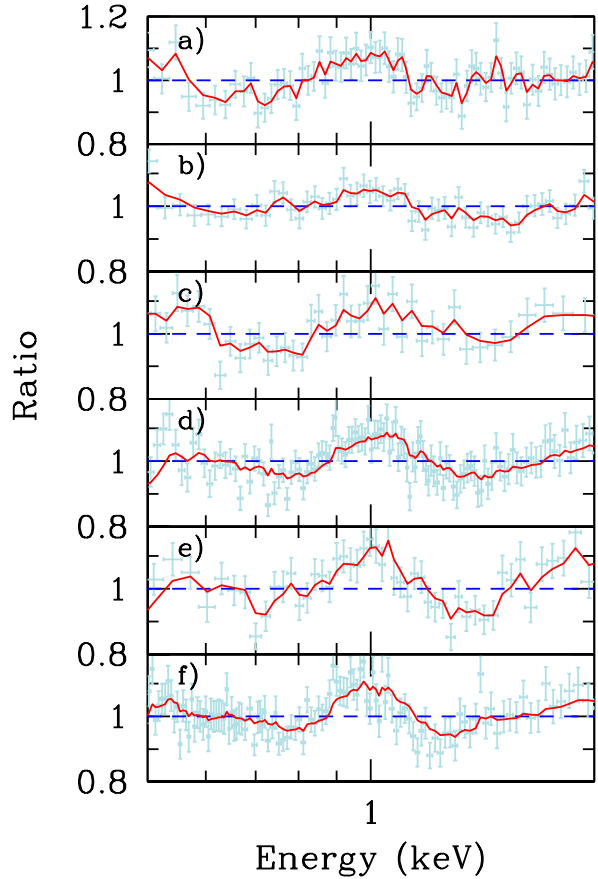}
\end{tabular}
\end{center}
\vspace{-0.2cm}
\caption{Plot showing the residuals to the best-fitting model of the continuum ({\sc tbabs*(diskbb+nthcomp)}: see M15)  for a) NGC 1313 X-1 (ObsID: 0405090101); b) Ho IX X-1 (ObsID: 0200980101); c) Ho II X-1 (ObsID: 0561580401); d) NGC 55 ULX-1 (ObsID: 0655050101); e) NGC 6946 X-1 (ObsID: 0691570101) and  f) NGC 5408 X-1 (ObsID: 0302900101). The spectral data (as a ratio to the best-fitting model) are plotted along with an exponentially smoothed function, the coefficient of which depends on the data quality and ranges from 0.4 in panel a) through to 0.8 in panel d) and e), and is 0.5 in the remaining sources. This indicates the overall shape and reinforces that the residuals are broad in the CCD spectra, important in light of the oversampling relative to the energy resolution of the instrument response. In all cases the residuals are significant and the energy at which they appear is remarkably similar across all five sources, implying a common origin.}
\label{fig:l}
\end{figure}

M14 have shown that the residuals are seen by both the MOS and PN (in the cases of NGC 5408 X-1 and NGC 6946 X-1), with the latter providing the vast majority of counts due the larger effective area. We therefore fit the EPIC PN spectra from our sample (across the 0.3-10~keV energy range) in {\sc xspec v12} (Arnaud 1996) with an absorbed continuum (with lower limit of n$_{\rm H}$ set at the Galactic column in the line-of-sight: Dickey \& Lockman 1990) of {\sc tbabs*(diskbb+nthcomp)} using the abundances of Wilms, Allen \& McCray (2000) and keeping the temperatures of the two components tied together (kT$_{\rm diskbb}$ = kT$_{\rm seed, nthcomp}$, for details see M15). With the exceptions of NGC~1313~X-2 and NGC~5204~X-1, we find strong evidence for residuals to the best-fitting model at soft energies in each source and plot the clearest examples of these in Figure 1 along with those for NGC~5408~X-1 and NGC~6946~X-1 for comparison. To confirm the statistical significance of these residuals, we perform a simple chi-squared fit to the data/model ratio between 0.5 and 2~keV with a straight line set equal to unity; in each case this is rejected (with a null hypothesis probability $\ll$ 0.05). We note that the features appear at approximately the same energies across the objects and crudely test this by fitting a simple model ({\sc constant + gauss + gauss}) to the full band-pass residuals. We find that the softer of the two features varies between 0.66 - 0.74~keV, consistent across sources within the individual 3-$\sigma$ error bounds. The harder of the features ranges from 1.23 - 1.46~keV with most of the values consistent within 3-$\sigma$ (with the exception of the lowest energy of these). As the residuals in each ULX appear at approximately the same energy (and with very similar apparent shapes), we take this to imply a common origin, which is important given that these sources span over an order of magnitude in (unabsorbed) luminosity from $\sim$1$\times$10$^{39}$ - 3$\times$10$^{40}$ erg s$^{-1}$ (M15). Finally we note that, although not obvious in the datasets analysed here, similar residuals have been seen in {\it Chandra} data of NGC 5204 X-1  when in a soft (and higher S/N) state (Roberts et al. 2006) and NGC 55 ULX-1 (Pintore et al. 2015); therefore these features are {\it independent} of detector, confirming that they are indeed genuine.

\section{Case study: NGC 1313 X-1}

In order to study the behaviour of the soft residuals and attempt to distinguish between possible origins, we require a source that evolves across a dynamic range in spectral hardness; as can be seen from Figure 8 and Tables 2 \& 3 of M15, the ideal source for such a study is NGC 1313 X-1. However, both the inferred (de-absorbed) hardness ratio and our ability to reliably characterise the residuals is sensitive to the modelling of the continuum. M15 show, via use of  covariance spectra (see Wilkinson \& Uttley 2009 and Uttley et al. 2014 for a review of the technique) that the ULX continuum emission originates from at least two components that cross at $\sim$1~keV. Mis-modelling of the continuum will therefore have an adverse effect on detecting atomic features imprinted between 0.7 and 2~keV, whilst an inability to separate out the components in the time-averaged spectrum can lead to misleading hardness ratios. To better address this problem we model each of the observations (using both PN and MOS data) using the continuum described above (and including a normalisation offset to account for differing responses, typically within $\sim$10 per cent of unity), finding that in five out of the available eleven observations we cannot distinguish between a solution where the soft component is intrinsically strong or weak. Middleton et al. (2015) show though use of the covariance spectrum, that the variability is entirely contained within the hard component but that a soft excess must still be present (see also Middleton et al. 2011) whilst the highest signal-to-noise data clearly show that the soft component dominates over the hard component below $\sim$1 keV. Whilst we can force the spectra for the five observations into what we believe to be a more appropriate deconvolution (namely by fixing the normalisation of the soft component - see M15), accurate modelling of the residuals requires the normalisations to be free to vary and so we exclude these observations from our subsequent analysis.     

\begin{table*}
\begin{center}
\begin{minipage}{185mm}
\bigskip
\caption{Spectral fitting results}
\begin{tabular}{|c|c|c|c|c|c|c|c}
  \hline
Model  & \multicolumn{4}{c}{\sc tbabs*gabs*gabs*(diskbb+nthcomp)} &  \multicolumn{3}{c}{\sc tbabs*(gauss+diskbb+nthcomp)} \\
\hline
 Obs. IDs & line strength & hardness &n$_{\rm H}$  & $\Delta\chi^{2}$ & line strength & hardness  &  $\Delta\chi^{2}$ \\
& (keV) &ratio & ($\times$10$^{22}$ cm$^{-2}$) & (1 d.o.f)  & (photons cm$^{-2}$ s$^{-1} \times$10$^{-5}$) & ratio & (1 d.o.f)\\ 
   \hline
 0205230401 & 0.17 $\pm$ 0.03 & 0.44 $\pm$ 0.07 & 0.37 $^{+0.04}_{-0.03}$ &  24.9 & 8.2 $^{+1.4}_{-1.3}$ & 1.81 $\pm$ 0.28 & 27.1\\
 0205230601 & 0.12 $\pm$ 0.03 & 1.00 $\pm$ 0.06 & 0.37 $\pm$ 0.04 & 18.7 & 5.9 $^{+1.2}_{-1.1}$ & 2.61 $\pm$ 0.12 & 21.4\\ 
 0405090101 & 0.07  $\pm$ 0.01 & 1.53 $\pm$ 0.05 & 0.31 $\pm$ 0.01 & 71.0 & 2.9  $\pm$ 0.3 & 2.55 $\pm$ 0.03 & 73.0\\
 0693850501,  0693851201 & 0.05  $\pm$ 0.01 & 1.73 $\pm$ 0.02 & 0.29 $\pm$ 0.01 & 84.0 & 2.8  $\pm$ 0.3 & 2.52 $\pm$ 0.03 & 90.3\\
 0205230301 & 0.01 $\pm$ 0.01 & 2.67 $\pm$ 0.04 & 0.27 $\pm$ 0.01 & 0.2  & $<$ 0.9 & 2.76 $\pm$ 0.07 & 0.0\\
\hline
\end{tabular}
Notes: Observations of NGC 1313 X-1, selected based on spectral deconvolutions that match expectation based on variability arguments (M15). In the case of observations 0693850501 and 0693851201 the hardness ratios (based on the ratio of the de-absorbed fluxes in the 1 - 10~keV to 0.3 - 1~keV bands) are similar and so we combine these into a single hardness bin. The values given in the table are the parameters of interest from the highly simplified models {\sc tbabs*gabs*gabs*(diskbb+nthcomp)} and {\sc tbabs*(gauss+diskbb+nthcomp)} with hardness ratios determined from use of the pseudo-model {\sc cflux} over the energy bands 1 - 10~keV and 0.3 - 1~keV respectively. The errors are quoted at 1-$\sigma$ (and in the case of the hardness ratios are propagated through from the mean errors on the flux in each band). The values of n$_{\rm H}$ are quoted for the former model only.  
\end{minipage} 

\end{center}
\end{table*}

\begin{figure}
\begin{center}
\begin{tabular}{l}
 \epsfxsize=8cm \epsfbox{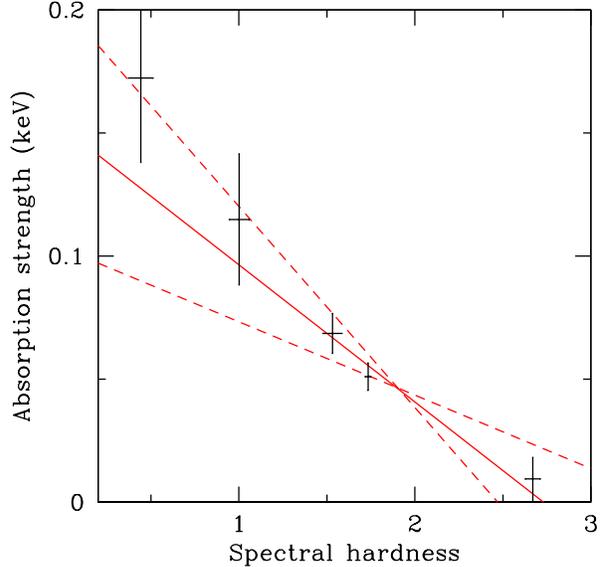}
\end{tabular}
\vspace{-0.2cm}
\end{center}
\caption{Absorption strength from the spectral fits ({\sc tbabs*gabs*gabs*(diskbb+nthcomp)}) plotted against spectral hardness (with 1-$\sigma$ error bars). The solid red line is the best-fitting relation (with a slope of -0.05 $\pm$ 0.01) and the dashed lines are the extremal range of relations (from determining the 3-$\sigma$ errors on the slope and intercept respectively).}
\label{fig:l}
\end{figure}

\begin{figure}
\begin{center}
\begin{tabular}{l}
 \epsfxsize=8cm \epsfbox{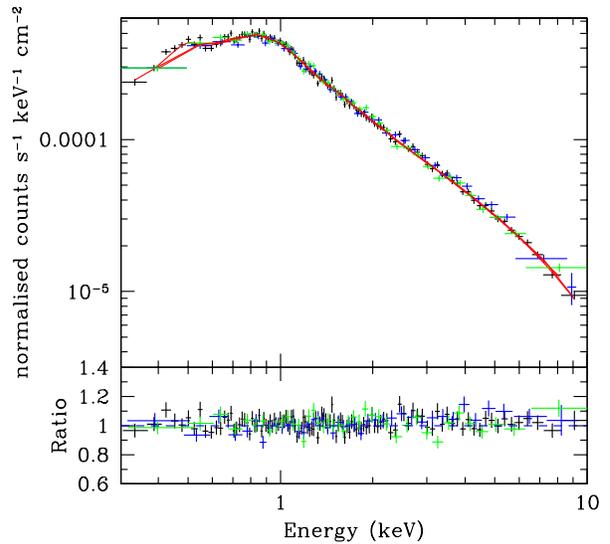}
\end{tabular}
\vspace{-0.2cm}
\end{center}
\caption{{\it Upper panel:} EPIC (PN: black, MOS: blue and green) data for NGC 1313 X-1, ObsID: 0405090101 and best-fitting folded model (in red: {\sc tbabs*gabs*gabs*(diskbb+nthcomp)}) plotted with the command {\sc setpl area} to avoid distorting effects of unfolding through the model (Nowak 2004). {\it Bottom panel:} the ratio of data/model from the above plot, showing an apparent lack of substantial deviations at soft energies (that are larger than the statistical error of the data).}
\label{fig:l}
\end{figure}

To provide a simple account of the residuals, we include two Gaussian absorption lines ({\sc gabs}) with the width ($\sigma$), line `strength' and centroid energies initially free to vary. In the case of {\sc gabs}, for low optical depth, the line strength is essentially the equivalent width in units of keV and we tie this between the two lines for the sake of simplicity. In {\sc xspec} the model is {\sc tbabs*gabs*gabs*(diskbb+nthcomp)} where we apply the absorption to the {\it entire} continuum emission. The improvement in chi-squared from including the lines is greatest for ObsID: 0405090101 ($\Delta\chi^{2}$ = 71 for 5 d.o.f.); we subsequently make the crucial assumption that the same lines are present in {\it every} dataset at the best-fitting energies (0.70 and 1.25~keV) and line-widths (0.11 and 0.23~keV) from the fit to ObsID: 0405090101, but at differing strengths (even when not detected significantly in a single observation). From the continuum model we determine initial hardness ratios from the ratio of the de-absorbed fluxes in the 1 - 10~keV and 0.3 - 1~keV bands; due to the similarities in hardness ratio we combine observations 0693850501 and 0693851201 into a single hardness bin which improves the statistics available for model fitting in this case. We apply the Gaussian absorption model with fixed line energies and widths (from the best-fitting model to ObsID: 0405090101) to the spectra, allowing the line strength (tied between lines) and the details of the continuum (including the neutral absorption column) to vary. The continuum model parameters are relatively unimportant for this analysis and for the sake of brevity we do not include them (instead we point the interested reader to M15 for a detailed discussion on such modelling and caveats). In Figure 2 we plot the resulting line strengths against the intrinsic (i.e. de-absorbed) hardness ratio, the latter determined by including the pseudo-model {\sc cflux} ({\sc tbabs*gabs*gabs*cflux*(diskbb+nthcomp)}) with 1-$\sigma$ errors on the latter determined from propagation of errors on the fluxes in the 1 - 10~keV and 0.3 - 1~keV bands. In the case of the bin containing observations 0693850501 and 0693851201, we obtain the weighted mean flux and hardness ratio (by tying the flux between observations). We determine the strength of the apparent anti-correlation between hardness and line strength using a Pearson product-moment correlation coefficient, finding an $r$ value (which takes values between -1 for a perfect anti-correlation to +1 for a perfect correlation) of -0.98. To test whether such a strong anti-correlation could result from random sampling, we perform a large series of Monte-Carlo simulations, selecting 5 Gaussian distributed points in absorption strength based on the observed mean and standard deviation (with the same hardness ratio values from the spectra). We find that, from 1 million simulations, only $\sim$1900 have an equal or better anti-correlation, implying a probability of $<$ 0.2 per cent that this is a random effect and as such the correlation appears genuine (at $>$ 3-$\sigma$).  By performing a least-squares fit with a linear model we find that the slope of the correlation is significantly negative ($<$ 0 at $\gg$ 3-$\sigma$) with a best-fitting index of -0.05 $\pm$ 0.01 (at 1-$\sigma$). 

As a further test, we determine the improvement in fit quality in each observation (or hardness bin respectively) when allowing the line width and energies to be free. The largest improvement in fit quality is for the bin containing observations 0693850501 and 0693851201 where we obtain a $\Delta\chi^{2}$ of 19 for 4 d.o.f. In this case, the values of the line widths and energies are all within 2-$\sigma$ of the frozen values and the line strength is within 1-$\sigma$ of the value reported in Table 1. Finally, we find that when the strengths are allowed to vary for each line independently, the resulting strengths are, in all but one case, consistent with each other and the best-fit tied value within 1-$\sigma$; in the remaining case (the bin containing Obs. IDs 0693850501 and 0693851201) the strengths are consistent within 3-$\sigma$.  

A possible weakness of our method is the assumption that the soft emission can be well described by a disc black-body. Should the underlying emission deviate substantially from this description, then it is possible the effect of neutral absorption could be degenerate with true changes at low energies, producing misleading hardness ratios. To demonstrate that this is unlikely to be an issue, we plot the raw data of ObsID: 0405090101 and best-fitting folded model (by using {\sc setplot area} to avoid distorting effects of folding through the model: Nowak 2004) together with the ratio of data to model in Figure 3; from inspection it would appear that any deviations at soft energies are only of a similar size to the statistical error on the data. However, we do note an apparent trend between neutral column density and the hardness ratio in the fits above; to illustrate this we plot the column density (with their 1-$\sigma$ errors) against the hardness ratio in Figure 4. In the case of the bin containing observations 0693850501 and 0693851201 we plot the mean column density and errors. We once again determine the Pearson product-moment correlation coefficient finding a value of -~0.93 and a chance probability via Monte-Carlo simulations of 1 per cent, demonstrating that this is not as strong as the anti-correlation between line strength and hardness ratio and we cannot rule this out as having a random origin. However, should this be a genuine correlation (with a slope of -0.04 $\pm$ 0.01), it could imply either a hidden degeneracy in our approach or might instead support a picture where a wind (probably associated with super-critical accretion) becomes increasingly dominant as the source softens (as seen to occur in observations of the ULX in NGC 55 - M15). Whilst a full understanding of such a scenario requires the putative wind to be physically modelled (and is beyond the scope of this paper), we speculate that should a wind be present and become stronger it may shield the surrounding material from the strong X-ray photon field, allowing some portion to recombine. Alternatively (or additionally) there may always be a neutral component associated with the outflow at large radii (where at some distance the material cools) and the column density may be linked to the mass loss which in turn is linked to the mass accretion rate and spectral shape (Poutanen et al. 2007). To further test for the impact of a potential degeneracy on our results, we fit all spectra simultaneously using the model above with the neutral absorption column tied between observations (see e.g. Miller et al. 2013). The resulting plot in Figure 5 (with a mean column density of [0.296 $\pm$ 0.001]$\times$10$^{22}$ cm$^{-2}$ at 1-$\sigma$) shows that the negative trend between absorption line strength ({\sc gabs}) versus hardness ratio remains (with a Pearson product-moment correlation coefficient  of -0.97 and a chance probability of 0.3 per cent), implying that the strength of the features is robust to a degeneracy in the column density.

\begin{figure}
\begin{center}
\begin{tabular}{l}
 \epsfxsize=8cm \epsfbox{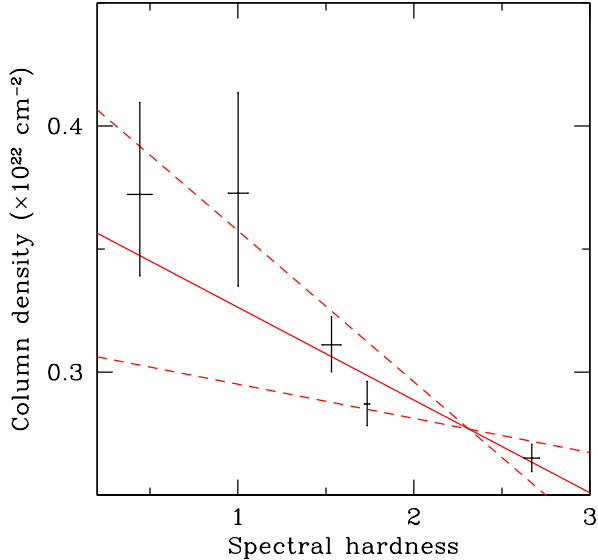}
\end{tabular}
\vspace{-0.2cm}
\end{center}
\caption{Neutral column density from the spectral fits ({\sc tbabs*gabs*gabs*(diskbb+nthcomp)}) plotted against spectral hardness (with 1-$\sigma$ error bars). The solid red line is the best-fitting relation (with a slope of -0.04 $\pm$ 0.01) and the dashed lines are the extremal range of relations (from determining the 3-$\sigma$ errors on the slope and intercept respectively).}
\label{fig:l}
\end{figure}

\begin{figure}
\begin{center}
\begin{tabular}{l}
\epsfxsize=8cm \epsfbox{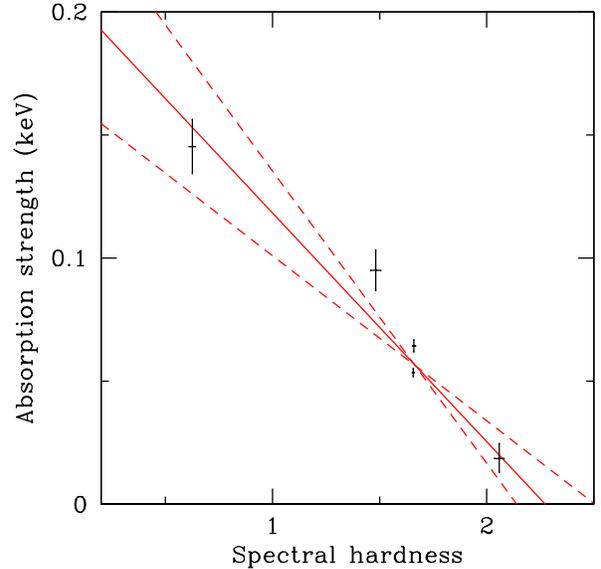}
\end{tabular}
\vspace{-0.2cm}
\end{center}
\caption{Absorption strength from the spectral fits ({\sc tbabs*gabs*gabs*(diskbb+nthcomp)}) plotted against spectral hardness (with 1-$\sigma$ error bars) when the neutral column density is tied between observations. The solid red line is the best-fitting relation (with a slope of -0.09$\pm$0.01) and the dashed lines are the extremal range of relations (from determining the 3-$\sigma$ errors on the slope and intercept respectively).}
\label{fig:l}
\end{figure}

As it is unclear whether the soft residuals are dominated by absorption along the line-of-sight or emission (which we assume to be isotropic), we repeat the above analysis but instead test for changes in the strength of an {\it emission} component by including a single Gaussian line in place of the two absorbers (i.e. {\sc tbabs*(gauss+diskbb+nthcomp)}: see e.g. Roberts et al. (2006) for a similar approach to fitting the {\it Chandra} spectrum of NGC 5204 X-1). In this case, the reported line strength is the total photon flux within the line. We once again find a highly significant improvement in fit quality for ObsID: 0405090101 ($\Delta\chi^{2}$ = 73 for 3 d.o.f.) and repeat our assumption that this component is present in all observations with the same best-fitting line energy (0.95~keV) and width (0.10~keV) as seen in  ObsID: 0405090101 but at varying strengths. We plot the best-fitting line strengths (i.e. the normalisation of the {\sc gauss} component as the line-width is fixed) against the hardness ratios in Figure 6, noting that the latter are now substantially changed (as we are using intrinsic hardness ratios which take into account absorption, including that by the {\sc gabs} components in the previous fits). Although a least-squares fit to the data yields a significant ($>$ 3-$\sigma$) negative slope of [-7.6 $\pm$1.7]$\times$10$^{-5}$ (with 1-$\sigma$ error), we find a correlation coefficient of only -0.82 with a random chance probability (from Monte-Carlo simulations) of 5 per cent. As can be seen from Figure 6 and Table 1, the correlation is significantly weakened by the fit to ObsID: 0205230601. It is plausible that the introduction of further emission components can lead to similar spectral degeneracies as discussed in our selection criteria and on closer inspection it would appear that the soft component is proportionally weaker in the fit to ObsID: 0205230601 than in the other spectra and thus the hardness ratio may be somewhat less reliable. We note that the changing line strength between observations provides further confirmation that the origin of the features cannot be in something as large-scale as galactic diffuse emission, which would not change on such relatively short timescales. 

For the spectrally hardest observation (ObsID: 0205230301) we have only been able to obtain an upper limit on the line strength when it is forced to be $\ge$ 0, i.e. a feature in emission. As the fitted value is at the limit of the parameter space, this can lead to a distortion in the error distribution (e.g. Kashyap et al. 2010). Although our least-squares fit to the data will be driven by the values with the smallest errors (i.e. those at moderate spectral hardness), to explore the likely impact of setting the lower limit on the line strength to zero, we allow the strength to be negative (i.e. an absorption feature). We find a best-fitting value of [0$^{+0.8}_{-0.9}$] $\times$10$^{-5}$ photons cm$^{-2}$ s$^{-1}$ (1-$\sigma$ error) . As the strength is unchanged and the errors remain relatively moderate in size, the distorting impact of setting the lower limit of the strength to be zero is expected to be minimal. 

\begin{figure}
\begin{center}
\begin{tabular}{l}
 \epsfxsize=8cm \epsfbox{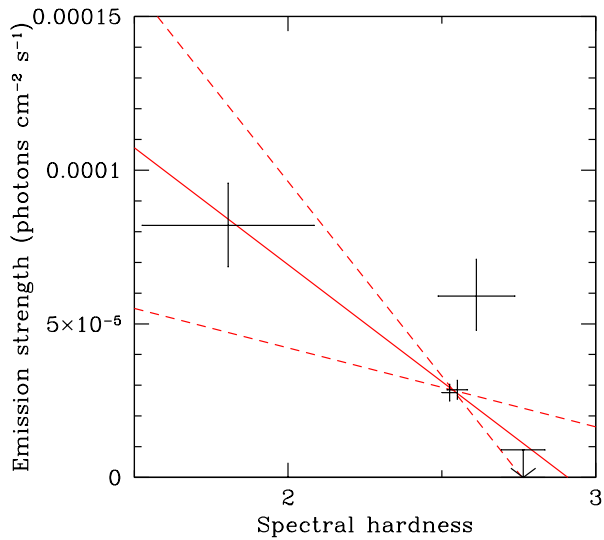}
\end{tabular}
\vspace{-0.2cm}
\end{center}
\caption{Emission line strength from the spectral fits ({\sc tbabs*(gauss+diskbb+nthcomp)}) plotted against spectral hardness (with 1-$\sigma$ error bars). The solid red line is the best-fitting relation (with a slope of [-7.6 $\pm$ 1.7]$\times$10$^{-5}$) and the dashed lines are the extremal range of relations (from determining the 3-$\sigma$ errors on the slope and intercept respectively).}
\label{fig:l}
\end{figure}

\vspace{-0.6cm}
\section{Discussion}

We have identified soft residuals to the continuum fits in several of the brightest ULXs, with their similar shapes implying that they are a common feature with a common origin. To explain their presence through an emission component requires extremely high luminosities which is inconsistent with emission from diffuse plasma associated with star formation (M14; Sutton et al. in prep). Such luminosities would also pose a problem for an origin in plasma emission from interactions between a disc wind and that from the secondary star (based on assumed ULX values if supercritical accretors). Instead we argue that the residuals are likely to be associated with atomic lines imprinted by either a wind close to the source (M14), in deposited material further from the accretion flow, via reflection (Fabian et al. 1989) or some combination of these possibilities. 

We have investigated the evolution of the soft residuals with underlying spectral shape for the case of NGC 1313 X-1 as the archival {\it XMM-Newton} data for this source are of high quality and show a substantial dynamic range in spectral hardness. Through a highly simplified spectral analysis we have identified a significant anti-correlation between absorption line strength and spectral hardness and a 
significantly negative trend (though an apparently weaker anti-correlation) between emission line strength and spectral hardness, allowing us to make comparisons to predictions for their origin:

{\bf Reflection:} In models describing the emission from ULXs, hard photons originating from either a hot inner disc (e.g. Poutanen et al. 2007) or from a large scale-height corona (e.g. Gladstone et al. 2009) illuminate optically thick material in the wind or disc. As a result, we should expect fluorescence features due to excited bound-free transitions and a Compton reflection `hump' due to down-scattering in the disc atmosphere (e.g. Fabian et al. 1989; and in specific relation to ULXs: Caballerro-Garcia \& Fabian 2010; Walton et al. 2011). As a result of {\it NuSTAR} observations which have extended the observational bandpass for even such relatively faint sources, the intrinsic continuum is now known to drop steeply above $\sim$6~keV, limiting the presence of a Compton hump (see e.g. Bachetti et al. 2013). Walton et al. (2013b) also place strict upper limits on the strength of reflected line emission around the expected energy of Fe K$_{\alpha}$; thus reflection is not thought to leave a strong imprint at high energies in ULX spectra, with the lack of line emission possibly the result of the illuminated material being very highly ionised, the reflected fraction very small or the Fe abundance being very low. Still the possibility remains that the features at soft energies could be the result of line emission due to reflection by material with a low Fe abundance but a high abundance of other elements (e.g. O and Ne - see M14). In such a scenario, where the hard illumination increases we should expect an increase in reflected emission and, where the change in spectral hardness is driven mostly by increases in the hard component (and not just the soft component falling away) as would generally appear to be the case (see Figures 5 \& 6 of M15), a {\it positive} correlation between emission line-strength and spectral hardness. Although the presence of an anti-correlation in emission line strength and spectral hardness (Figure 6), is admittedly not as strong as for absorption line strength, we can at least rule out a positive trend in the data. As a complicating factor we note that this ignores changes in the ionisation state as a result of changing the incident luminosity. Although the ionisation state {\it must} change, the range in luminosities covered by these observations is $<$ 1 dex (M15). Should we be observing reflection features it therefore seems unlikely that the material would become so highly ionised as the source brightens without requiring density or geometry changes.


It is also important to consider changes in the reflected fraction that would result from changing the solid angle to the illuminating flux, as this will have a marked effect on the strength of any fluorescence features. In a model invoking an inner corona, an increase in hard emission may imply a larger covering fraction to the incident photons, a larger reflected fraction and a positive correlation once again. In the model invoking a conical wind, the increase in hardness is thought to be associated with an increase in mass accretion rate and a smaller opening angle of the wind cone leading to increased geometrical beaming (King 2009). As the inner radius of the wind is not expected to be sensitive to such changes (Poutanen et al. 2007), the corollary is that the reflected fraction should increase giving a positive correlation once again.



{\bf Absorption in a wind:} In the super-critical model of ULXs (Shakura \& Sunyaev 1973; Poutanen 2007) the wind is launched from an advection-dominated disc, creating an evacuated wind cone, the inner surface of which is expected to be highly ionised for typical densities and ULX luminosities. At small inclinations to the line-of-sight, the spectrum at high energies is dominated by scattering/geometrical beaming of flux from the inner disc out of the wind cone whilst the soft emission originates in the disc/wind itself (Poutanen et al. 2007). As the wind cone is evacuated, only the soft emission will be substantially affected by absorption by the wind material, with the features (in reality we should expect both absorption and emission) broadened due to velocity dispersion along the line-of-sight (M14; Pinto et al. in prep). At more moderate inclinations, the hard emission is likely to be down-scattered through the wind whilst the soft emission will typically dominate; as both components are seen through the wind, absorption should affect {\it both} components at these inclinations. 

Should the mass accretion rate through the disc increase (possibly due to longer timescale trends imprinted at large radii), the cone of the wind should close, with the hard emission from the inner disc becoming increasingly beamed (King 2009). For observers at low inclinations, the hard component will get stronger and, as this emission is not absorbed, it will dilute the strength of the absorption features which overlap in energy (M15). At more moderate inclinations, the beaming should not have a significant impact on the spectrum and the strength of the absorption should be relatively independent of flux changes. Therefore for those sources identified as being viewed into the wind cone we have the expectation of an anti-correlation with spectral hardness (and flatter when at moderate inclinations). This would appear to match observation (Figure 2), where we have discovered a highly significant anti-correlation between absorption line strength and spectral hardness. However, we note a probable degeneracy in this solution; an increase in mass accretion rate will likely lead to a larger deposition of material by the wind (see Poutanen et al. 2007), a larger column density and therefore stronger features. The emission from the continuum spectral components and amount of geometrical beaming will also change (with the observational effect dependent on the inclination: M15) and so the spectral hardness will change as a result. Studying this degeneracy in full requires careful modelling of the wind and the radiative transfer which is beyond the scope of this work. However, we can make the simple prediction that at moderate inclinations the soft emission will likely increase whilst the hard emission will be increasingly beamed away and Compton down-scattered in the line-of-sight. Therefore, for sources seen at moderate inclinations, an increase in mass accretion rate could also lead to an anti-correlation in spectral hardness and line strength; distinguishing between these possibilities requires careful modelling (and consideration of the time-dependent properties: M15). 

{\bf Atomic features from distant material:} A further possible explanation for the residuals is the presence of atomic features (in absorption and/or emission) imprinted onto the spectrum by photo-ionised material that is further from the accretion flow  -  deposited by either a wind or jet (Cseh et al. 2014) - though not so far that the density and optical depth become ineffectual (and not so far away from the source that the strength of the features cannot change on inter-observational timescales). 

In the case where the optical depth of the putative surrounding material is not expected to change on inter-observational timescales we should expect the strength of the features to remain unchanged with flux (although the details of the lines, e.g. ionisation state, may change should the flux change considerably) resulting in an approximately flat relationship with spectral hardness. This does not appear to be the case given the negative correlations of Figures 2 \& 6, although should the material not be isotropically distributed but instead preferentially located in the equatorial plane, with winds blocking the harder emission, then an anti-correlation would be expected (as with the case of absorption in a wind, described above). This would naturally invoke the presence of winds and likely some measure of geometrical beaming, however, the features would not necessarily be broad but could instead be a blend of narrow lines. Should mass deposition by a wind into the surroundings be varying - which in turn would affect the strength of the features by varying the density and optical depth - then we would return to the scenario described above with only a simple prediction for a correlation with mass deposition and spectral state (and which requires careful modelling to be fully understood).

Should a persistent ballistic-type jet be present in this source (such as that seen in SS433 and a potential analogue  discovered in Ho~II X-1: Cseh et al. 2014) it may instead be responsible for introducing a mass flux into the local environment of the ULX leading to changes in line strength. However, unless the spectrum is somehow tied to the mass flux into the jet, then we should not expect a correlation. Whilst this cannot presently be ruled out, an analogy to a ballistic type of ejection would already invoke high mass accretion rates (and in the case of SS433, super-critical rates of accretion), where winds are ubiquitous and may well dominate mass loss from the accretion flow (Neilsen, Remillard \& Lee 2011; Ponti et al. 2012) and are well known to leave an imprint in absorption and emission at moderate inclinations (e.g. Ponti et al. 2012). Additionally, assuming that the features are at least partially seen in absorption, we would require that the entire ULX population be viewed within a relatively small range of inclinations (of order the jet opening angle) which seems unlikely given the variety in spectral shape and variability properties (M15). A possible solution is to invoke system precession so that the jet deposits material across a large solid angle (as seen in SS433: Fabrika et al. 2004) although this would still require that any mass flux be tied to the spectral emission of the central source.

\vspace{-0.6cm}
\section{Conclusion}

The behaviour of the soft X-ray spectral residuals in ULXs may provide an important lever-arm for diagnosing the nature of the luminous accretion flow. Whilst in the past they have been ascribed to thermal plasma emission associated with star formation (a reasonable assumption given that many ULXs are found in star forming regions, e.g. Gao et al. 2003), on closer inspection (e.g. M14; Sutton et al. in prep) this association does not hold as the star formation rate would predict far lower luminosities {\it across the whole star-forming region} than required. This is especially constraining for those with low star formation rates (e.g. NGC 1313: see the discussion in Bachetti et al. 2013). A natural origin may be shocks between an outflowing wind and that of the stellar companion, however, taking what we assume to be appropriate values for super-critical accretion (based on those for SS433 and radiatively driven outflows), this also predicts luminosities one to two orders of magnitude below that required. Other possible explanations for the features include reflection from neutral to partially-ionised, optically thick material in a wind or disc, absorption (and some amount of emission) in an outflowing wind (potentially broadened and analogous to those in BALQSOs) or absorption/emission in photo-ionised material close to the ULX (or some combination of these). 

Fitting simple models to the EPIC spectra of NGC 1313 X-1 suggests a significant negative slope between the strength of the features and spectral hardness, arguing against an origin in reflection but instead favouring either absorption in a wind or features imprinted by material further out but shielded from the hard X-ray flux by an inner wind (such that the features are diluted when the hard component becomes stronger). Notable caveats associated with this claim include our assumption that the features are not inherently transient (which seems reasonable in light of the persistent nature of the source) and that the modelling is extremely simple and does not yet include a self-consistent treatment of absorption and emission.

NGC 1313 X-1 shows relatively unusual evolution in its time-averaged spectrum, appearing at times extremely soft and at others extremely hard (Pintore \& Zampieri 2012; M15). This has proven decisive in allowing us to detect features when soft (as with NGC 5408 X-1 or NGC 6946 X-1: M14) and trace their evolution in apparent strength when the source becomes harder. There are several possibilities to explain the spectral evolution of the source within the super-critical model (Poutanen et al. 2007), depending on inclination angle and mass accretion rate (M15). Should the source be viewed at moderate inclinations it is plausible that the source is at higher mass accretion rates when softest; if the source is viewed face-on it may instead be at lower mass accretion rates (so that the hard component is less geometrically beamed) or equally, the system might precess (probably due to a slaved disc in a similar manner to SS433, e.g. Fabrika et al. 2004) so that occasionally we see into the wind, producing a softer spectrum. M15 favour this latter scenario (based on joint spectral-timing arguments) and is consistent with our finding that the neutral column density appears to also anti-correlate with spectral hardness (with only a moderate random chance probability) which would occur when more material enters our line of sight, (although this would also occur as a result of the mass accretion rate increasing). Further modelling (and use of advanced timing analysis tools) to test the predictions is necessary to conclusively break this degeneracy but is beyond the scope of this paper and for the purposes of identifying the origin of the residuals is unnecessary - a wind is favoured irrespective of the details.


A corollary of our selection of a wind model to explain the behaviour of this source is that some amount of geometrical beaming must be invoked. Isotropic emission arguments associated with the optical nebula of the ULX Ho II X-1 (Pakull \& Mirioni 2002; Kaaret, Zezas \& Ward 2004, and which also shows similar spectral residuals - Figure 1) would argue against a large amount of beaming in this and presumably similar sources. However, the optical line luminosity limits combined with the variability of the X-ray source may still allow for beaming factors of a few (typically less than an order of magnitude) which is compatible with a model for super-critical accretion and large scale-height winds. The super-critical wind model also has the advantage of explaining the lack of strong reflection features in the hardest sources (e.g. Walton et al. 2013b) due to the inner face of the wind cone being highly ionised (although at some distance from the central source - and should the density be high enough - some weak reflection features should still be produced). 

We note that a negative correlation between absorption strength and spectral hardness is also seen in Galactic BHBs (Ponti et al. 2012) as a result of spectral evolution and radiative/MHD powering of winds from the disc surface (e.g. Neilsen \& Homan 2012). Whilst at first glance this may match the observed behaviour of NGC 1313 X-1 and potentially imply considerably larger black hole masses and sub-Eddington accretion states, the lack of reflection features would be at odds with such a picture. In addition, the spectral changes seen in BHBs are clearly not a good match to those seen in ULXs (e.g. M15).

It remains as yet unclear whether the features are intrinsically broad (due to velocity dispersion in a wind) or are a complex of narrow lines e.g. from material further from the accretion flow. Whilst the nature of the lines could be unambiguously determined from dispersion grating spectra (RGS onboard {\it XMM-Newton} or LETG/HETG onboard {\it Chandra}), the typical count rates generally preclude a meaningful analysis (although we will address this in a parallel work: Pinto et al. in prep). {\it Astro-H} will provide the combination of throughput and energy resolution needed to address this, whilst dedicated campaigns with ESA's upcoming {\it ATHENA} mission will allow high quality tests for the behaviour of these important features across the wider ULX population. 

\vspace{-0.5cm}
\section{Acknowledgements}

The authors thank the anonymous referee for their suggestions and Simon Vaughan for helpful discussion. MJM appreciates support from ERC grant 340442, DJW is supported by ORAU under the NASA Postdoctoral Program at the NASA Jet Propulsion Laboratory. TPR was funded as part of the
STFC consolidated grant ST/L00075X/1. This work is based on observations obtained
with {\it XMM-Newton}, an ESA science mission with instruments and
contributions directly funded by ESA Member States and NASA.

\label{lastpage}

\end{document}